\documentclass{nws}

\usepackage{amsmath}
\graphicspath{ {Figures/} }

\begin{document}

\title[Network Science]
      {Symmetry in cancer networks identified : Proposal for multi-cancer biomarkers}

 \author[P. Shinde] 
 		{Pramod Shinde \\       
		   Discipline of Biosciences and Biomedical Engineering, Indian Institute of Technology Indore, Indore 453552\\
         ({\it e-mail}: \texttt{pramodshinde119@gmail.com})}
         
  \author[L. Marrec]
        {Lo\"ic Marrec\\
         Sorbonne Universit\'e, CNRS, Laboratoire Jean Perrin (UMR 8237), Paris F-75005 France\\
         ({\it e-mail}: \texttt{loic.marrec@sorbonne-universite.fr})}
                          
  \author[A. Rai]
        {Aparna Rai\\
        Discipline of Biosciences and Biomedical Engineering, Indian Institute of Technology Indore, Indore 453552\\
         ({\it e-mail}: \texttt{raiaparna13@gmail.com})}
                 
   \author[A. Yadav]
      {Alok Yadav\\
         Complex Systems Lab, Discipline of Physics, Indian Institute of Technology Indore, Indore 453552\\
       ({\it e-mail}: \texttt{physicistalok@gmail.com})}
       
   \author[R. Kumar]
        {Rajesh Kumar\\
           Discipline of Physics, Indian Institute of Technology Indore, Indore 453552\\
         ({\it e-mail}: \texttt{rajeshkumar@iiti.ac.in})}

   \author[M. Ivanchenko]
        {Mikhail Ivanchenko\\
        Department of Applied Mathematics and Centre of Bioinformatics, Lobachevsky State University of Nizhny 	 
        Novgorod, Nizhny Novgorod, Russia\\
         ({\it e-mail}: \texttt{mikhail.ivanchenko@itmm.unn.ru})} 
          
  \author[A. Zaikin]
        {Alexey Zaikin\\
        Department of Mathematics and Institute for Womens Health, University College London, London, WC1E 6BT, UK\\ 		and\\
        Department of Pediatrics, Faculty of Pediatrics, Sechenov University, Moscow, Russia\\
         ({\it e-mail}: \texttt{alexey.zaikin@ucl.ac.uk})}

  \author[S. Jalan]
        {Sarika Jalan*\\
           Complex Systems Lab, Discipline of Physics, Indian Institute of Technology Indore, Indore 453552 \\  
		   Discipline of Biosciences and Biomedical Engineering, Indian Institute of Technology Indore, Indore 453552\\ and \\
		   Lobachevsky University, Gagarin avenue 23, Nizhny Novgorod, 603950, Russia.\\
         ({\it e-mail}: \texttt{sarikajalan9@gmail.com})}

\jdate{\today}
\pubyear{2019}
\pagerange{\pageref{firstpage}--\pageref{lastpage}}
\doi{S}

\newtheorem{lemma}{Lemma}[section]


\maketitle
\begin{abstract}
One of the most challenging problems in biomedicine and genomics is the identification of disease biomarkers. 
In this study, proteomics data from seven major cancers were used to construct two weighted protein-protein interaction (PPI) networks {\it i.e.,} one for the normal and another for the cancer conditions. 
We developed rigorous, yet mathematically simple, methodology based on the degeneracy at -1 eigenvalues to identify structural symmetry or motif structures in network.  
Utilising eigenvectors corresponding to degenerate eigenvalues in the weighted adjacency matrix, we identified structural symmetry in underlying weighted PPI networks constructed using seven cancer data.
Functional assessment of proteins forming these structural symmetry exhibited the property of cancer hallmarks. 
Survival analysis refined further this protein list proposing {\it BMI}, {\it MAPK11}, {\it DDIT4}, {\it CDKN2A}, and {\it FYN} as putative multi-cancer biomarkers. 
The combined framework of networks and spectral graph theory developed here can be applied to identify symmetrical patterns in other disease networks to predict proteins as potential disease biomarkers.\\
{\bf keywords:} Cancer networks, Eigenvalue analysis, Graph symmetry, Biomarkers.
\end{abstract}

\tableofcontents

\section{Introduction}
Each cancer tissue comprises a heterogeneous and multi-factorial milieu that varies in cytology, physiology, signaling mechanisms, cell regulation, control mechanisms and response to therapy. 
Both the existence of genetic diversity among tumors of same cancer and the surprising amount of similarity among different cancers have been reported \cite{Stratton}.
Interestingly, similarities among different cancers have widely been observed in cell proliferation rate, cell-cell interactions, metastatic potential and sensitivity to therapy.
Therefore, the abundance of similarity in different cancers has allowed us to consider cancer as a single system in the present study. 

Furthermore, many biological processes can be modeled as graphs composed of interactions among numerous cellular and molecular components \cite{Bara, PS3}.
Probing a complex system in network or graph theory framework allows understanding a phenomenon or system's behavior by collecting information of all its constituents rather than focusing to a smaller part with apparent relation with a phenomenon \cite{SarkarPhysics}.
Network studies have been providing global understanding to corresponding biological processes and functional interactions \cite{Cai, Dosztanyi, PS2, PS4}.
Important outcomes were that different types of biological networks exhibit network features such as complexity, robustness and hierarchical behavior \cite{Bara}. 
Cancer network based studies helped to predict protein function, genotype-phenotype relationships between cancer proteins, a combined effect of DNA, RNA, protein modifications on overall cancer development and impact of mutations in altering molecular pathways \cite{Yixuan, Lage}.
These cancer network studies have been successful in developing drug strategies as well as in finding important cancer pathways, e.g., mTOR signaling, p53 pathway, MAPK, and PI3K signaling pathways \cite{Yildirim, Chiang}.
However, these investigations have focused mainly on structural positions of proteins or pathways in underlying networks.

Moreover, biological networks have been found to possess abundant symmetrical patterns \cite{Wang}.
Symmetrical structures such as motifs have been heavily investigated for their relevance of biological processes \cite{PS3, PS4, Christofk, Cheng}.
Motifs are complete subgraphs, represent building blocks of many biological networks and these structures have been reported in cellular networks of diverse organisms from bacteria to humans, suggesting motif structures are highly conserved in evolution \cite{Cheng}. 
Functional failure of such local structures can have substantial global impacts \cite{Milo}.
For instance, a group of tumor suppressor genes forming onco-modules recently identified whereas oncogenic mutations in these modules altered the pan-cancer metabolic landscape \cite{Cubuk}.
In this work, we focused on spectral (eigenvalues) properties of the network adjacency matrix for unraveling symmetrical patterns and corresponding proteins in the underlying network.
The importance and uses of spectra of adjacency matrices have been well characterised in various model networks as well as real-world networks  \cite{SarkarC, Rai3, Agrawal}.

Degeneracy in graph spectra has contributed significantly in our understanding of structural and dynamical properties of corresponding graphs \cite{Yadav, Mieghem}.
The driving force behind the investigation of origin and implication of degenerate eigenvalues is that biological networks constructed using empirical data show very high degeneracy, particularly at 0 and $-1$ eigenvalues, than corresponding random networks \cite{PS1, LM, Rai3}.
Indeed, these degenerate eigenvalues have been shown to exist due to an outcome of the complete and the partial node duplication \cite{Yadav} which is akin of the fundamental process in the evolution-related with gene duplication and diversification process \cite{PS1, Teichmann}. 
As part of the cell cycle, particularly during replication of the genome, seldom another copy of a gene is synthesized. 
Immediately after this gene duplication event, both the original gene and the new identical copy of the gene have the same DNA sequence, so both interact with the same set of molecular partners. 
Consequently, as these genes are guided for their particular functions, each of the molecular partners that interacted with the ancestor gains a new interaction \cite{Teichmann}. 
Similarly in cancer genomes,  clonal duplication and proliferation are achieved by DNA mutations \cite{Furlong}, mainly using somatic copy number alterations \cite{Zack}. 
The gene duplication and diversification process play a crucial role in the growth, adaption, evolution, and subsistence of the biological system \cite{Teichmann}. 
Though degeneracy at $-1$ eigenvalue can be related to specific structures in a network, the origin and implication of such structural patterns are not that obvious.
Herein, we focused on symmetrical patterns corresponding to $-1$ degenerate eigenvalues and devised methodology to identify such essential network symmetrical structures.

In this work, we first provided a methodology to identify an origin and implications of eigenvalue degeneracy in weighted networks. 
Second, we applied this technique to find structural patterns corresponding to degenerate eigenvalue in weighted multi-cancer PPI network.
Network structures linking to $-1$ degeneracy provided a framework for identification of proteins corresponding to underlying local patterns.
The functional assessment further deduced that these proteins corresponding to $-1$ eigenvalue degeneracy have the property of cancer hallmarks.
With survival analysis, we predicted cancer proteins {\it i.e.,} {\it BMI}, {\it  MAPK11}, {\it DDIT4}, {\it CDKN2A}, and {\it FYN} as putative multi-cancer proteins.\\

\begin{figure*}[htbp]
{\includegraphics[width=13cm, height=6.5cm]{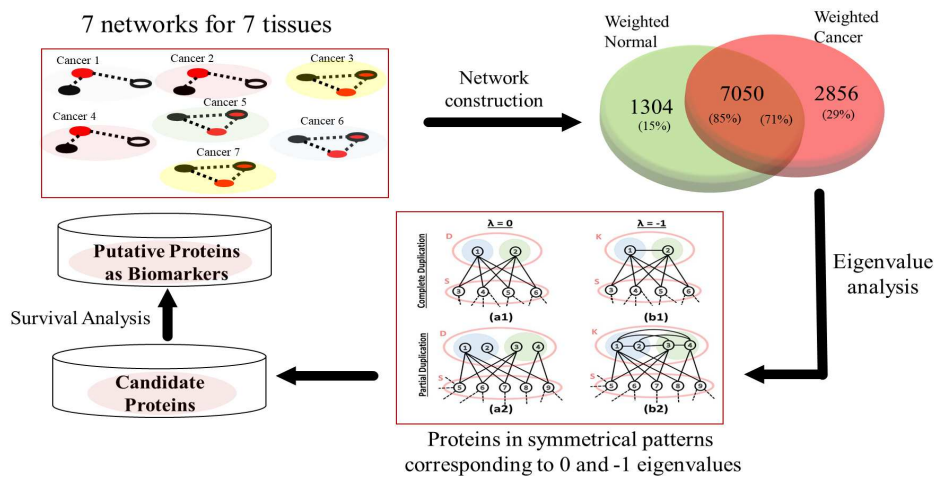}}
\caption{{\bf Work-flow diagram depicting network construction, eigenvalue analysis, and identification and characterisation of multi-cancer biomarkers.} }
\label{FigA}
\end{figure*}

\section{Data and methods}
\subsection{Dataset sources}
We constituted our multi-cancer PPI network using proteomics data from morphologically seven different cancers such as Breast, Cervical, Colon, Lung, Oral, Ovarian and Prostate.
For straight-forward comparison of the proteome in a diseased state, we also retrieved PPIs in corresponding healthy tissue states.
We termed healthy tissues as `normal' and cancer tissues as `disease'.
In this way, we have 14 datasets {\it viz,} seven for healthy tissues and seven for disease tissues.
PPIs in a healthy and the corresponding disease tissues were designated on the basis of their occurrence in the normal or the diseased tissue, respectively. 
PPI data mining was broadly divided into two steps, {\it i.e.,} (I) retrieval of protein names pertaining to a particular tissue, and then (II) retrieval of PPI of corresponding to proteins identified in step I.

In the first step, protein-name data mining was independently performed on each tissue (Figure \ref{Fig1}(a) and \ref{Fig1}(b)). 
Here, it is to be noted that we used the text-mining approach to map proteins for a particular dataset.
Similarly, other approaches such as proteins corresponding to highly expressed genes can be utilized to map proteins.
Also, it should be noted that our protein-name data mining was entirely based on the information available in secondary bioinformatics databases which are already curated and largely followed data sources {\it viz,} UniProtKB and GeneBank databases.
Protein-name data mining was performed using different search words, and accordingly, protein-names were destined to a particular dataset.
For example, if a protein entry in the UniProtKB database has been related to the information of oral cancer tissue, we marked that protein entry as a member of oral cancer dataset.
The details of search words (queries) used for protein-name mining from these databases is given in Supplementary Materials. 
Additionally, we explored other resources to enrich our protein-name collection. 
Swiss-2DPage (https: //world- 2dpage. expasy.org/swiss -2dpage/) and Cervical cancer database (CCDB) (http:// crdd.osdd. net/raghava /ccdb/) for cervical tissues, ACTREC Oral Cancer Database (http:// www. actrec. gov.in/ OCDB/ index.htm) and Head and Neck Oral Cancer Database (http:// gyanxet.com/ hno.htm) for oral cancer, ATCC cell line database (https: //www. atcc.org/) and Cancer Cell Line Encyclopedia (https://portals. broadinstitute. org/ccle) for all considered cancers.  
The detailed list of proteins for both healthy tissues and the corresponding cancer tissues collected from various literature and data archives can be found at \cite{dataToUse}.
In the second step, once all the proteins for seven different tissues for the normal and disease states were collected, leading to fourteen datasets, the interacting partners of these proteins were retrieved from the STRING database version 9.189 \cite{String}.
We used the default parameters in STRING database while retrieving PPI's. 
An interaction between a pair of proteins was considered if there exists a direct ({\it i.e.,} physical), indirect ({\it i.e.,} functional) or both relation between them. 
Direct PPI interactions are straight-forwardly measured between protein pairs, whereas indirect PPI interactions are identified using the information of one or more bridging molecules.

In this way, we have seven datasets for the normal and seven datasets for the corresponding disease states.
The detailed information of these fourteen datasets representing PPIs among all the fourteen tissues can be found at \cite{dataToUse}.

\subsection{Weighted multi-cancer PPI network construction}
In a PPI network, vertices represent proteins and edges represent interactions between the proteins.
We overlaid PPIs derived from seven tissues in two datasets {\it i.e.,} (1) normal and (2) disease, separately to construct two weighted PPI networks.
Weights were assigned based on edge overlapping {\it viz,} the number of times an interaction is found in a set of cancers (schematic is provided in Fig. \ref{Fig1}(c)). 
For instance, if an interaction between two nodes $k$ and $l$ found in colon and breast cancer, and was absent in other cancers that would yield a weight, $w_{kl} = 2$. 
Consequently, each element in the adjacency matrix has value ranging from 1 (min) to 7 (max).
If an interaction existed in all seven cancers, the corresponding weight entry in the adjacency matrix would be 7 and if an interaction existed in only one cancer in the adjacency matrix, the weight entry would be 1.
The weighted adjacency matrix can be given as:
\begin{equation}
W_{\mathrm {ij}} = \begin{cases} w_{\mathrm {ij}}~~\mbox{if } i \sim j \\
					0 ~~ \mbox{otherwise} \end{cases}
\label{adj_wei}
\end{equation}
where $w_{\mathrm{ij}}$ was the number of times $i$ interacted with $j$.
In such manner, we considered two interaction matrices, one for healthy state and another one for the disease state.

\subsection{Various structural measures of a network}
The most basic structural parameter of a network would be the degree of a node ($k_i$), which can be defined as a number of weighted edges connected to the node $i$ ($k_i=\sum_j w_{ij}$).
Further, the clustering coefficient ($C$) can be defined as a ratio of the number of interactions a neighbor of a particular node is having and the possible number of connections the neighbors can have among themselves.
For a weighted network, $C$ can be defined as the geometric average of the subgraph edge weights, 
$C_{i} = \frac{1}{k_{i}k_{i-1}}\sum_{j,k}(\hat{w_{ij}}\hat{w_{jk}}\hat{w_{ik}})^{\frac{1}{3}}$ \cite{Saramaki}.
The edge weights $\hat{w}$ were normalized by the maximum weight in the network $\hat{w} = \frac{{w}}{max(w)}$.
Betweenness centrality \cite{Brandes} of a node $i$ defined as the sum of the fraction of all-pairs shortest paths that were passing through $i$, such that 
$\beta{c}(i) =\sum_{s,t \in V} \frac{\sigma(s, t|i)}{\sigma(s, t)}$ where $V$ was the set of nodes, $\sigma(s, t)$ was the number of shortest $(s, t)$-paths, and $\sigma(s, t|i)$ was the number of those paths passing through some node $i$ other than $s, t$. 

\begin{figure*}[t]
\centerline{\includegraphics[width= 11cm]{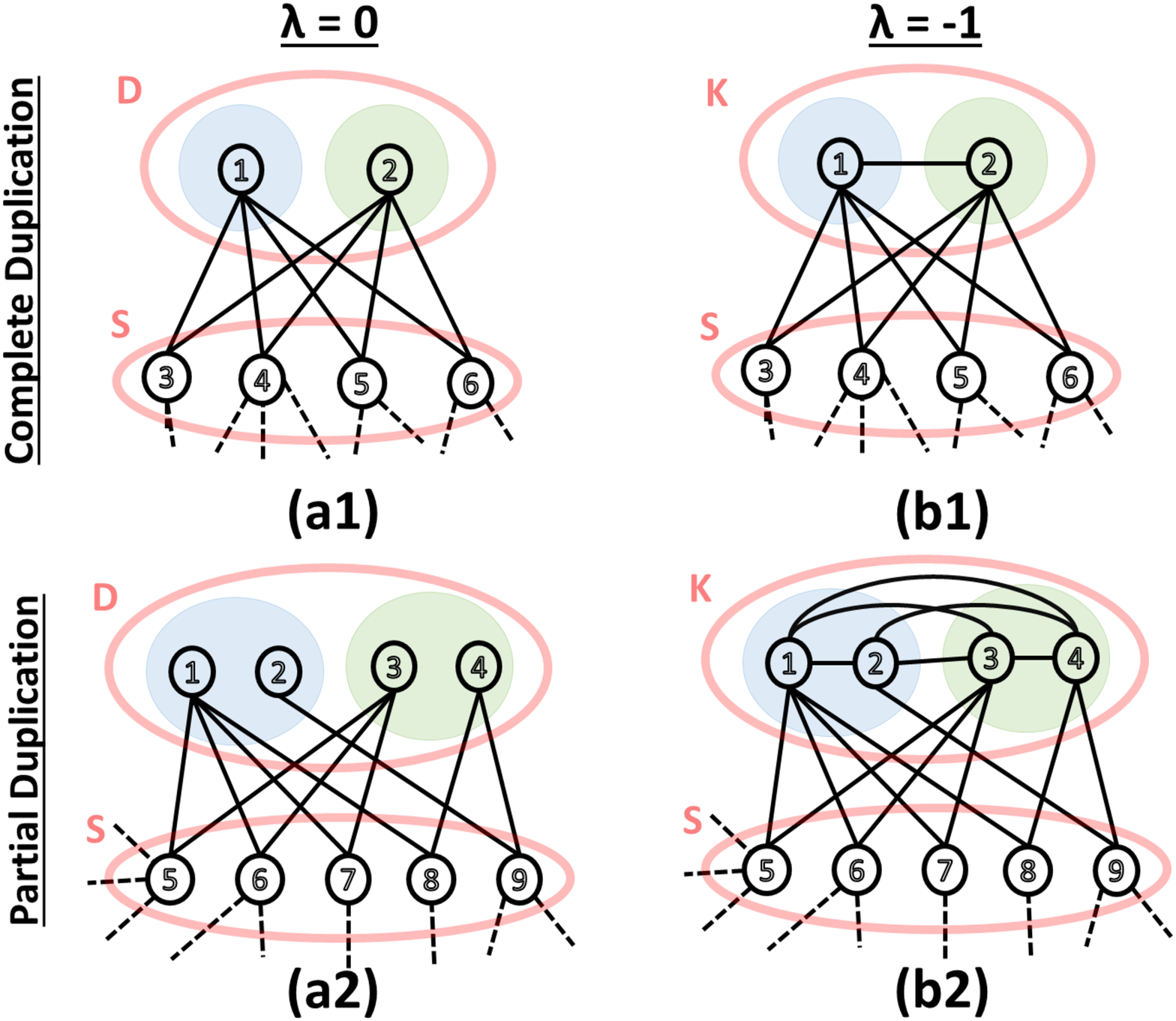}}
\caption{{\bf Structures leading to $0$ and $-1$ eigenvalue degeneracy in weighted cancer PPI network.} 
(a1 and a2) for $\lambda = 0$ and (b1 and b2) for $\lambda = -1$. Note that to fulfil partial duplication condition will require more number of nodes than those required to fulfil complete duplication. It can be seen that there exists only one difference between the structures corresponding to 0 and -1 degeneracy. In the case of 0 eigenvalue degeneracy, there is no interaction between the nodes being completely ({\it i.e.,} 1 and 2) or partially duplicated ({\it i.e.,} 1,2,3 and 4) whereas there exists interaction between them in the case of -1  eigenvalue degeneracy.}
\label{Struct}
\end{figure*}

\subsection{Theoretical framework: relating structural symmetry and degenerate eigenvalues in weighted networks}
We considered finite undirected and weighted graphs defined by $G=\{V,E\}$ with $V$ is the node set, and $E$ is the edge set such as $|V|=N$ and $|E|= N_c$. 
Eigenvalues $\lambda_1 \leq \lambda_2 \leq ... \leq \lambda_N$ were obtained by computing the roots of the characteristic polynomial of ${\bf W}$.
Note that the eigenvalues were real because ${\bf W}$ was symmetric. 
The associated eigenvectors \textbf{v}$_1$, \textbf{v}$_2$ ,..., \textbf{v}$_N$ satisfied the eigen-equation ${\bf W} \textbf{v}_{i}=\lambda_{i}\textbf{v}_{i}$ with $i=1,2,...,N$.

All the origin and implications of $0$ eigenvalue degeneracy in networks spectra has been already well characterized \cite{Yadav}. 
Briefly, the spectrum of a matrix of size $N$ and rank $r$ should encompass $0$ eigenvalue with multiplicity $N-r$ \cite{Rank}. 
There exist three conditions which would lead to the lowering of the rank of a matrix: $\mbox{(i) }R_i=(0 0 \ldots 0)$ a row with only zero-entries.
$\mbox{(ii) }R_i=R_j$ at least two rows are equal (Figure \ref{Struct} \texttt{(a1)}). 
$\mbox{(iii) }\sum_i a_i R_i=\sum_j b_j R_j\mbox{ with } a_i,b_j \in{\Bbb R}$ two or more rows together are equal to some other rows (Figure \ref{Struct} \texttt{(a2)}).
We would not consider the condition (i) which was related to the isolated nodes in ${\bf W}$.
Additional information regarding $0$ degeneracy can be found in earlier studies \cite{Yadav}.

As precribed in \cite{LM}, it was possible to reduce the computation of $x$-eigenvalue of ${\bf W}$ matrix to the $0$-eigenvalue of (${\bf W}-x{\bf I}$) matrix. 
Now, lets understand the occurrence of $x$-eigenvalue degeneracy in weighted networks and see when should conditions (ii) and (iii) get fulfilled in (${\bf W} -x{\bf I}$) which was written as follows: 
\begin{equation}
{\bf W} -x{\bf I} = 
\begin{pmatrix}
-x & w_{12} & \cdots & w_{1N} \\
w_{12} & -x & \cdots & w_{2N} \\
\vdots  & \vdots  & \ddots & \vdots  \\
w_{1N} & w_{2N} & \cdots & -x 
\end{pmatrix}
\label{W}
\end{equation}
Note that we considered graphs without self-loops. 
The condition (ii) can met if and only if: $w_{ik}=w_{jk} \mbox{ with }k=1,2,...,N$.
In the particular case $R_1=R_2$, the previous equation becomes:
\begin{equation}
  \begin{cases} w_{12}=-x  \\
	w_{1k}=w_{2k} \mbox{ with }k=3,4,...,N 
  \end{cases}
\end{equation}

Specifically, the case of $-1$ eigenvalue can be related to $K*S$ structures (Figure 2).
In these structures, all the nodes of $K$ were interlinked with the same weight $w_K$.
In addition, all the nodes of $K$ are connected to the same set of neighbours, $S$ having identical weight, $w_{Si}$ (Figure \ref{Struct} \texttt{(b1)}). 
For (${\bf W}-x{\bf I}$) matrix, we would get the relation $w_{K} = -x$. 
By this way, $D*S$ structure can be seen as a particular case of $K*S$ with $w_K=0$ (Figure \ref{Struct} \texttt{(a2)}). 
This has highlighted one of the most interesting aspects of degeneracy in weighted graphs. 
The condition (ii) would always give a $K*S$ sub-graph and the only difference was based on the weight of edges. 
More particularly, the weight of edges in $K$ was directly related to the eigenvalue to which it contributes. 
However, because of this supplementary constraint, we would expect a lower degeneracy resulting from the condition (ii) in weighted networks as compared to unweighted networks. 
Indeed, it was sure that most of the $K*S$ structures observed in unweighted networks would not fulfill this constraint if the weights were taken into account.

Further, to simplify condition (iii), we considered the particular case $R_1 + R_2 = R_3 + R_4$, which gave:
 \begin{equation}
 w_{1k} + w_{2k} = w_{3k} + w_{4k} \mbox{ with } k = 1,2,...,N
 \end{equation}
 The last equation can be developed as a system:
 \begin{equation}
\begin{cases} w_{12} - x = w_{13} + w_{14} = w_{23} + w_{24}  \\
w_{34} - x = w_{13} + w_{23} = w_{14} + w_{24} \\
w_{1k} + w_{2k} = w_{3k} + w_{4k} \mbox{ with }k=5,6,...,N
 \end{cases}
 \end{equation}

Contrary to the condition (ii), the condition (iii) did not shed light on a typical structure. 
Indeed, since $w_{ij}$ can take any real value, the number of possible solutions were high and so it should be difficult to find a general solution to the previous equation. 
This was due to the fact that condition (iii) can result from a linear combination of rows. 
By this way, we would expect a degeneracy resulting from the condition (iii) at more eigenvalues in weighted graphs than in the case of unweighted graphs and so, contrary to the case of condition (ii), we would observe more different structures which were not brought in to light by degeneracy in unweighted networks. 
Here we would limit ourselves to provide an example of graph that could verify $R_1 = R_2 + R_3$ in $W$ and (${\bf W} + {\bf I}$) (see Figure \ref{Struct} \texttt{(b2)}). 
 
So far, we focused on finding structures behind occurrence of eigenvalue degeneracy. 
The next question was: {\it could we identify the nodes involved in such structures ?} The answer was yes since it had been shown that it's possible by using the eigenvectors associated to degenerate eigenvalues \cite{LM}. 
More particularly, the components of these eigenvectors verify the following relation:
 \begin{equation}
\left\{
    \begin{array}{ll}
	\sum_{i\in K_p} v_{i}=0 \mbox{ with } v_{i}\neq0 \mbox{ and }p=1,2,...,n_{K*S}  \\
	v_{j \in V\setminus \{K_1\cup K_2 \cup ... \cup K_{n_{K*S}}\}}=0
    \end{array}
\right. 
\label{KS} 
\end{equation}
for nodes belonging to $K*S$ structures, where $n_{K*S}$ denoted the number of such sub-graphs in the whole network.

Similarly, for the nodes which has belonging to a sub-graph verifying the condition (iii) in ${\bf W}-x{\bf I}$, one has the relation:
\begin{equation}
 \left\{
    \begin{array}{ll}
	\sum_{i\in (L.C)_p} v_{i}=0 \mbox{ with } v_{i}\neq0  \mbox{ and }p=1,2,...,n_{L.C}\\
    v_{j \in V\setminus \{(L.C)_1\cup (L.C)_2 \cup ... \cup (L.C)_{n_{L.C}}\}}=0 
    \end{array}
\right.
\label{LC} 
\end{equation}
where $L.C$ and $n_{L.C}$ were the linear combinations and the number of linear combinations, respectively. 
Thanks to these relations, one could identify easily the nodes contributing to degenerate eigenvalue.
One way to handle this issue was to consider the matrix (${\bf W} -x{\bf I}$) and to search for each $R_i=R_j$, which was computationally doable. 
Then, we could consider one of the eigenvectors associated to $x$ eigenvalue and identified all the non-null entries. 
These should not obey $R_i=R_j$ to belong necessarily to a linear combination.
\\

\subsection{Gene enrichment and survival analysis}
We used genes from significant signatures {\it i.e,} corresponding to -1 eigenvalue degeneracy as an input into STRING \cite{String}, Panther \cite{Panther}, and MSigDB \cite{MSigDB} gene ontology platforms.
Further, we measured the correlation between each gene activity and patient survival outcomes using Cox proportional risks group hazards models available with SurvExpress biomarker validation tool \cite{SurvExpress} for TCGA cancer 
gene expression data (Supplementary materials). 
TCGA database provides a catalogue genetic mutations responsible for over 20,000 primary cancer and matched normal samples spanning 33 cancer types.
In particular, we inserted details of Gene name(s) and Tissue of interest in the data-fields given at SurvExpress homepage. Then, we have chosen the TCGA database for data retrieval and performed Biomarker Cox survival analysis using default parameters.

\begin{figure*}[htbp]
{\includegraphics[width=13cm, height=11cm]{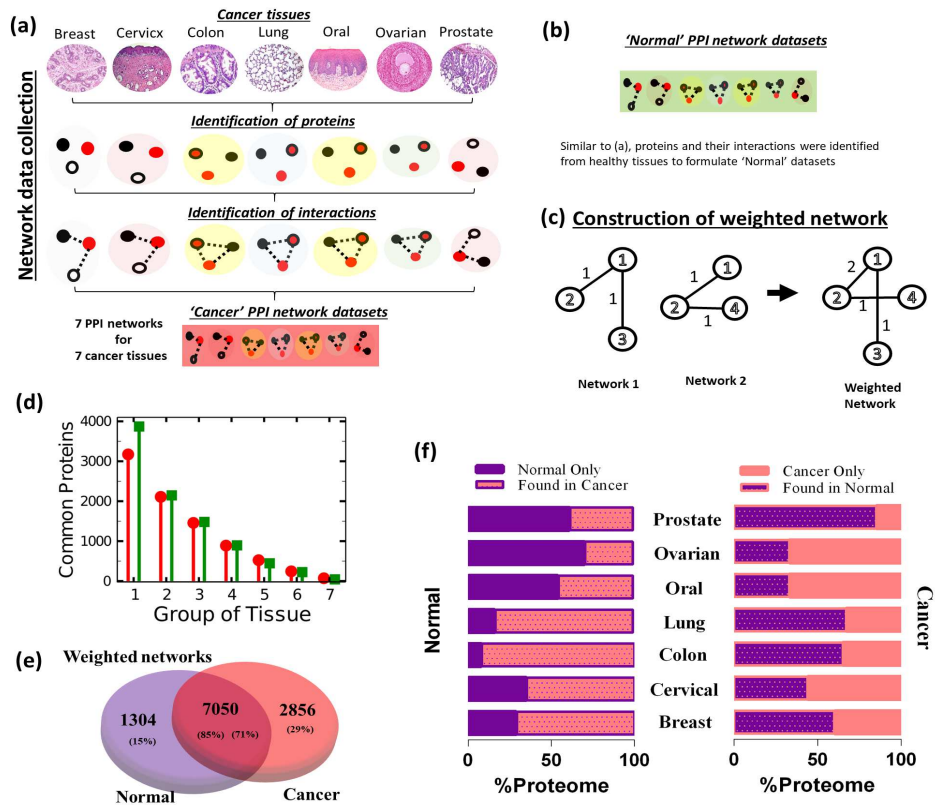}}
\caption{{\bf Construction and analysis of weighted multi-cancer PPI network.} (a and b) Schematic diagram showing overall network data collection from healthy and cancer tissues, (c) Schematic diagram illustrates the construction of weighted network where edges from two networks (in actual seven) constitute one weighted network. (d) Common proteins found between group tissues in normal and cancer datasets. (e) Venn diagram shows overlap between the number of proteins found in healthy and cancer tissues. (f) shows the number of times a protein can be found in a particular group of tissue. For instance, if a protein is present in breast and oral cancer, it is said to present in a group of two tissues.  }
\label{Fig1}
\end{figure*}

\begin{table}
\caption{{\bf Properties of (un-) weighted normal and disease PPI networks.} Here, $N$ means number of nodes, $N_{C}$ means number of connections, $\langle K \rangle$ means average degree, $k_{max}$ means degree of the hub node, $\lambda_{-1}$ means number of minus one eigenvalues, $\lambda_{-1}^{nodes} (ii)$ and $\lambda_{-1}^{nodes} (iii)$ represent number of proteins linked to degeneracy corresponding to $\lambda_{-1}$ with condition $ii$ and $iii$, respectively. $^w$ stands for weighted network and $^{unw}$ stands for unweighted network.}
\begin{tabular}{cccccccccc}    \hline
Network &$N$ \hspace{30mm}& $N_c$ \hspace{30mm}& $\langle k^{unw} \rangle$ \hspace{20mm}& $\langle k^{w} \rangle$ \hspace{20mm}& $k_{max}^{unw}$ \hspace{10mm}&$k_{max}^{w}$ \hspace{10mm}&$\lambda_{-1}$  \hspace{20mm}&$\lambda_{-1}^{nodes} (ii)$ \hspace{20mm}&$\lambda_{-1}^{nodes} (iii)$ \\ \hline
Normal  &9946 \hspace{40mm}&105491 \hspace{20mm}&21 &32 \hspace{20mm}&636 \hspace{20mm}&1565 	&23 &39	&0 \\  \noalign{\vspace {.5cm}}
Disease &8354 \hspace{40mm}&102701 \hspace{20mm}&25 &35 \hspace{20mm}&877 \hspace{20mm}&1409 	&20 &35	&4 \\\hline 
\end{tabular} 
\label{tab1}
\end{table}

\section{Results}
\subsection{Analysis of normal and cancer datasets}
Before we would present results based on the analysis of weighted networks, we outlined few observations about number of proteins in the healthy and cancer tissues by considering all the cancers as a single unit.
Existence of common proteins in both the normal and corresponding cancer states have suggested that their common aetiology and common functions which are essential for cell survival and growth.
We found that more than 65\% ($\pm$ 23\%) proteins in the normal tissues were found in corresponding cancer tissues when we considered each tissue separately (Figure \ref{Fig1}(C)).
However, there were as much as 85\% of proteins in normal tissues were found in cancer tissues when we took all the normal tissues as a single unit (Figure \ref{Fig1}(D)).
It suggested that a large portion of proteome of healthy tissues have role in some or other cancer related activities. 
Similarly, more than 51\% ($\pm$16\%) proteins from individual cancer tissues were found in normal tissues when we considered each tissue separately (Figure \ref{Fig1}(C)).
Interestingly, there were as much as 71\% of cancer proteins were also present in normal tissues when we considered all cancer tissues as a single unit (Figure \ref{Fig1}(D)).
It would be institutive to have a higher proteome overlap when different tissues were considered as one unit but it was interesting to note that cumulative cancer tissue proteome has less overlap than cumulative normal tissue proteome. 

\subsection{Importance of nodes based on structural properties of networks}
We constructed two types of networks for both the disease and the normal datasets: (1) unweighted networks which was constructed based on the presence and the absence of interactions between proteins, and (2) weighted networks where weights were assigned to an interaction based on the number of times an interaction was repeated in the combined list (Figure \ref{Fig1}(A)).
First, we examined the structural properties of these networks.
We found that $\langle k \rangle$ was higher in the disease networks than that of normal networks (for both the weighted and the unweighted cases) suggesting that cancer proteins have more affinity to interact among themselves.
Further, we found that the highest degree nodes ($k_{max}$) in the unweighted and the weighted multi-cancer networks were different (Supplemetary materials).
The hub protein in unweighted cancer network was {\it UBC} ($k = 877$) whose pathway function is translation regulation whereas the hub protein in weighted multi-cancer network was {\it CACNB2} ($k = 1565$).
Additionally, the top 10 degree proteins in weighted multi-cancer network were also among pathway regulators (Supplemetary materials).
This observation lie in accordance of known fact that the regulatory proteins were high degree proteins in PPI networks \cite{Fox}.
Second, weighted multi-cancer network has {\it CACNB2} and {\it BRD7} ($k$ = 1524) as two high degree proteins in which {\it CACNB2} has the role among CCR5 pathway in macrophages and PEDF induced signaling (http://www.proteinatlas.org/ENSG00000165995-CACNB2/cancer) and {\it BRD7} has TP53 activity \cite{Yu}.
It was interesting to note that though {\it CACNB2} and {\it BRD7} perform essential cancer activities \cite{Yu}, they are yet to get thoroughly investigated for drug related activities in cancer.
Nevertheless, weighting scheme have provided identification of another set of nodes which were vital for cellular processes in cancers under investigation.

\begin{figure*}[htbp]
\centerline{\includegraphics[width = 14cm, height = 9cm]{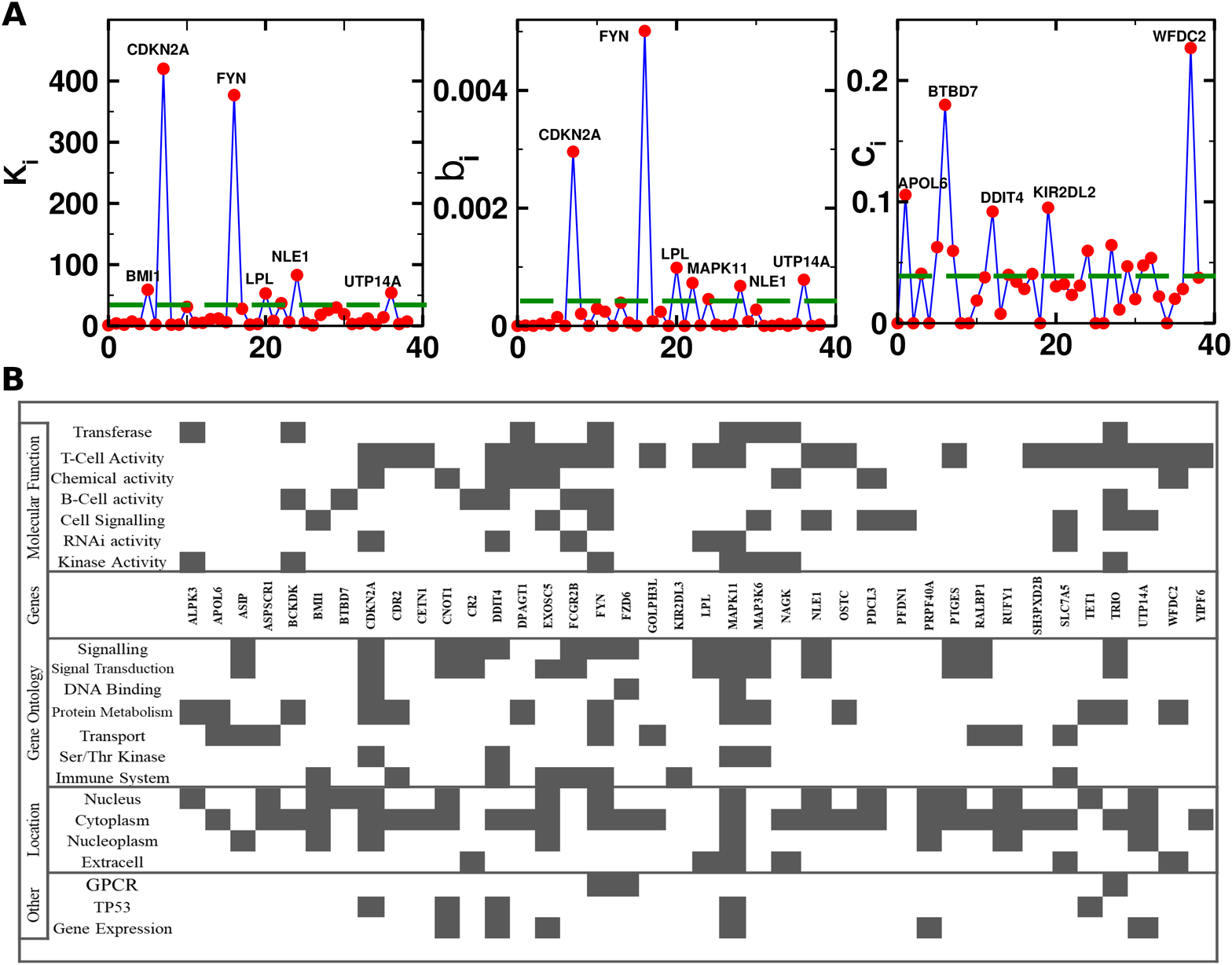}}
\caption{{\bf Network structural properties and functional assessment of 39 cancer proteins identified using network symmetry.} (A) Network structural properties such as degree ($k$), betweeness centrality ($\beta_C$) and clustering coefficient ($C$) for 39 proteins are displayed where proteins showing high value of particular property are highlighted. Horizontal line (green) shows the average value.(B) Gene functional assessment is categories into four groups {\it i.e.,} molecular function, gene ontology, location and other. }
\label{Fig2}
\end{figure*}

\subsection{Importance of nodes corresponding to degeneracy in weighted cancer network}
Next, we discussed to the prime focus of the current study by screening nodes forming structural patterns corresponding to -1 eigenvalues ($\lambda_{-1}$) in weighted multi-cancer PPI network and further noted down their biological significances and network properties. 
There were 39 proteins corresponding to -1 eigenvalues (summarized in Supplemetary materials Table 1).
These proteins, except {\it APOL6} and {\it BTBD7}, have been reported to be related to more than one tumors.
Each of 39 proteins posses one or more property of cancer's hallmarks \cite{Hanahan} as they have participated in cell signaling, signal transduction, transport etc (Figure \ref{Fig2}(B)).
Additionally, they exhibited essential bio-physical and bio-chemical activities such as enzymatic (kinase, transferase), immunological (B-cell, T-cell) and molecular (RNAi, signalling) activities.
Few of them were also found to perform activities at multiple cellular locations such as cell nucleus, cytoplasm, nucleoplasm and extracellular matrices which was biologically more relevant in performing specific biological activities related to cellular communications (Figure \ref{Fig2}(B)). 
It was interesting to report that these 39 proteins did not take any significant structural position in global-level weighted multi-cancer PPI network. 
Therefore, they were not detectable at global-level network using various measures such as node degree, clustering coefficient and betweenness centrality (Figure \ref{Fig2} and Supplemetary materials) available in network literature to identify them as structurally important nodes.

Since it was known that disease biomarkers would tend to have higher degree and connectivity in comparison to non-disease genes because of higher values of gene expressions \cite{WinterD}, among these 39 proteins we first focused on those proteins which have degree higher than average degree of 39 nodes.
Second, we focused on proteins having $\beta_{c}$ higher than $\langle \beta_{c} \rangle$ of 39 proteins.
Betweenness centrality measures the extent to which a node lies on paths between other vertices.
The removal of nodes with high betweenness centrality from the network will most disrupt communications between other nodes because they lie on the largest number of paths \cite{Barh}.
It's known that high degree nodes have high $\beta_c$ value.
However, there were interesting reports where moderate degree nodes have high $\beta_c$ and these nodes were proposed to be selected as effective cancer targets \cite{Barh}.
Lastly, we noted down proteins with higher $C$ values than $\langle C \rangle$ of 39 proteins. 
Clustering coefficient demonstrates cluster forming ability of nodes or how well a node is connected among its direct neighbours \cite{BarabasiReview}. 
Interestingly, above five nodes were among top 10 nodes with high $C$ value in our weighted multi-cancer PPI network.
Overall, we short-listed 12 significant candidate proteins by refining the list of 39 proteins with respect to network properties.

\begin{figure*}[t]
\centerline{\includegraphics[width=12cm, height=8cm]{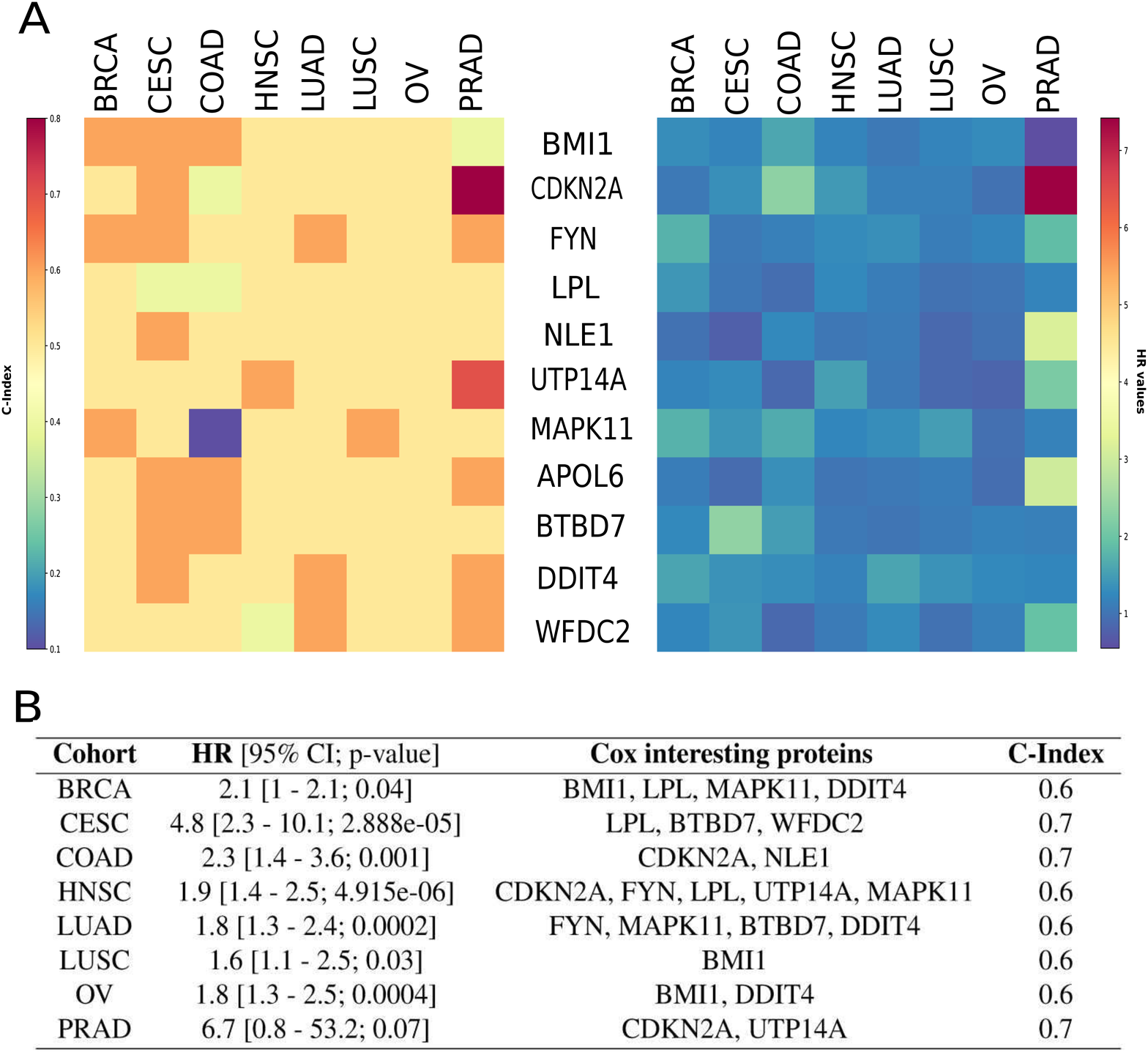}}
\caption{{\bf Comparison of biomarkers using overall survival analysis.} The listed 11 proteins are distinguished proteins among 39 proteins. (A) Single protein analysis. C-index and HR values of 11 candidate proteins in cancer cohorts are shown. For example, {\it CDKN2A} showed significant HR value in COAD (HR 2.3 [95\% 1.2 - 4.4), p=0.01]). It is to note that we select two cancer cohorts for lung cancer {\it i.e.,} LUAD and LUSC. (B) Multi-protein analysis where group of 11 candidate proteins are analysed against each cancer cohort. Kaplan-Meier curves for both single protein and multi-protein analysis are given in Supplementary materials.}
\label{Fig5}
\end{figure*}

\subsection{Survival Analysis} 
It's essential to study function of a protein with its role in patient survival to devise cancer biomarker \cite{Brock}.
To achieve this, we assessed whether the selected 12 candidate proteins were also associated with the overall survival (OS) in different cancers.
Out of 12 proteins, we did not find data of patient survival for {\it KIR2DL2}. 
Therefore, we selected 11 proteins for further analysis.
First, we performed OS with multi-protein (11 proteins) to understand the role of degeneracy in eigenvalues which arises due to underlying symmetry in interaction in each cancer (Figure S13).
Risk hazard ratio (HR) measures an effect of an intervention on an outcome of interest over time whereas C-index measures and compares the discriminative power of a risk prediction model. 
HR value 1 means lack of association, a hazard ratio greater than 1 suggests an increased risk, and a hazard ratio below 1 suggests a smaller risk.
HR value was found be more than 1.5 for each cohort in which Cervical squamous cell carcinoma (CESC) and Prostate adenocarcinoma (PRAD) displayed significantly high value (Figure \ref{Fig5}).
Secondly, we carried out the single-protein analysis which identified most significantly associated proteins with OS in each cancer, independently (Figures S3-S12).
We identified five proteins as putative multi-cancer biomarkers {\it i.e.,} {\it BMI}, {\it MAPK11}, {\it DDIT4}, {\it CDKN2A}, and {\it FYN}.
These proteins have HR value more than average HR value for atleast 3 cohorts as well as they occurred in atleast 3 cohorts in multi-protein analysis.

Furthermore, we compared the differential expression of BMI, MAPK11, DDIT4, CDKN2A, and FYN proteins between low-risk and high-risk patient groups in each cancer cohort. 
The definition of low-risk and high-risk patient groups were taken from SurvExpress, which generates two equal groups of patients cohort using the prognostic index \cite{SurvExpress}.
In particular, we identified {\it BMI1} which was epigenetic regulator and it promoted oncogenesis with DNA damage response \cite{Nacerddine}.
Our OS analysis found that {\it BMI1} has increased levels of gene expression in high-risk groups in seven cohorts (expect in HNSC) (Supplemetary materials).
Interestingly, we found that {\it CDKN2A} was absent in Breast cancer adenocarcinoma (BRCA) cohort and it has decreased expression in high risk patients (except Colon adenocarcinoma (COAD) and PRAD).
We also found that {\it CDKN2A} showed significant HR value in COAD (HR 2.3 [95\% 1.2 - 4.4), p=0.01]) and almost significant HR value in PRAD (HR 7.4 [95\% 0.9 - 59.1, p=0.06]) cohorts.
Further, we found the increased level of {\it MAPK11} expression in high-risk patients of BRCA (Supplemetary materials) which was supported by the fact that {\it MAPK11} was highly expressed in the metastatic breast cancer \cite{MAPK11}.
In a way, it was possible to correlate increased {\it MAPK11} expression in high risks patients with high HR value. 
Interestingly, we found that {\it DDIT4} has high HR values in all fast growing cancers (Supplemetary materials).
{\it DDIT4} is considered to be a driver in the aggressiveness of cancer cells because of its apoptotic activity.
{\it DDIT4} is induced by a variety of stress conditions and inhibit mTORC1 pathway \cite{Pinto}.
{\it FYN} is known to be up-regulated in human prostate cancer and has role in cancer progression and metastasis \cite{Elias}.
Overall, our survival analysis predicted {\it i.e.,} {\it BMI}, {\it MAPK11}, {\it DDIT4}, {\it CDKN2A}, and {\it FYN} as putative multi-cancer proteins which could effectively stratify low and high-risk cancer patients.

\section{Discussion}

Our analysis focused on weighted PPI network constructed based on the number of times a particular interaction among couple of proteins present in seven most prevalent and morphologically different cancers. 
Most of earlier works were typically node-centric where as we adopted a holistic approach excessively exploiting significance of functional interactions among different cancer tissues. 
In a way, our method provided an ultimate scope for identification of a protein set that would not have over-represented otherwise.
The observation of cumulative cancer tissue has lesser protein overlap than normal tissues suggested the possibility of diverse cancer related activities within and across cancer tissues. 
{We also found that hub proteins in unweighted and weighted networks were completely different.
In unweighted network, {\it CACN2} and {\it BRD7} were turned out as two high degree proteins in which both of them were largely unexplored for their therapeutic use. 
In other case, {\it UBC} was identified as hub protein in unweighted network which was very well known for cancer related activities. 
All this implicated the significance of weighing scheme to identify another set of nodes which were vital. 
 
Significance of this approach laid in the identification of simple and precise, yet fundamental, symmetrical structures of underlying network through -1 eigenvalues.
Biological network motifs drive very specific functions depending on the needs of the cell.
Though many efforts have been devoted to identify network motifs to capture particular local functionality within a biological network, still scope persist for efficient method development.
Our method identified symmetrical structures in the underlying weighted PPI networks and picked up proteins forming these essential network structures as candidate proteins.  
These symmetrical structures were based on degeneracy in $-1$ eigenvalues. 

Symmetrical structures presented here depict groups of proteins having a structural equivalence in a network.
Degeneracy at $-1$ eigenvalue essentially detects pairs of nodes which are not only connected to exactly the same other nodes (similar to $0$ eigenvalue degeneracy), but also connected to one another forming motifs structures.
Each network motif or a complete subgraph can operate as an elementary circuit with a well-defined function, which is integrated within a larger network and has a role in performing the required information processing. 
Such recurring elementary circuits have already been emphasised within varieties of biological networks including cancer networks \cite{Hanahan}.
In general, degeneracy in cancer can be understood in terms of independent adaptation of each cancer gene arising due to natural selection \cite{Hanahan}.
In present context, we used network structures corresponding to $-1$ eigenvalues as a measure of degeneracy in network graph.
This structural phenomenon is very interesting since many proteins are essentially backups for others, and can perform similar functions if one is knocked out or not functional at a particular phase of the cell cycle. For example, when a eukaryotic cell is exposed to ionizing radiation, a group of RAD52 proteins attends as a backup pathway operating independently in place of DNA dependent protein kinase \cite{Perrault}. In another example, the presence of two distinct pathways of glycoproteins and non-glycoproteins exist in mammalian cells for translocation of misfolded proteins from the endoplasm reticulum (ER) to the cytosol \cite{Ushioda}. First one is functional in non-stress condition and later is functional in ER stress \cite{Ushioda}. One more interesting example is that a significant number of cancerous mutations found to fall at structurally equivalent positions within the proteins catalytic core, particularly in kinases \cite{Dixit}. These structurally equivalent positions are also termed as mutational hotspots \cite{Dixit}.

The identified 39 proteins corresponding to patterns linked to $-1$ eigenvalue degeneracy did not take any significant structural position in weighted multi-cancer PPI network and hence they were not detectable using various measures such as node degree, clustering coefficient and betweenness centrality.
However, these proteins should have profound effects on information processing in the protein-protein interactions since their position in a network arised due to underlying symmetry among interactions.
In addition, the list of 39 proteins showed important biological roles given by gene enrichment analysis.
Essentially, because these 39 proteins were not hub proteins, their removal would have little impact on the overall statistics of the network which was essential to rid of false positive outcomes.

Further, we short-listed 12 significant candidate proteins by refining the list of 39 proteins with respect to network properties.
Finally, we convinced with five putative proteins which displayed high HR values in both single- and multi-protein analysis.  
Also, these five proteins displayed very specific roles in group of cancers in survival analysis.
The current study demonstrated that the spectral graph theory framework is a powerful concept and tool for revealing important structural patterns in network.
Utilizing networks, cancer biomarkers were identified considering their stands in pathways and cycles instead of mere higher values of network features alone. 

\section{Conclusion}
{The current study was focused on the importance of interactions between proteins participating among various cancer tissues.
Two main objectives were currently pursued: first, the glance at functional interactions among all cancers as single unit, which permitted us to look at all cancer related processes under one data framework;  and second, the use of network theory and spectral graph theory as a means to identify important causative agents for multi-cancer diagnosis and therapy.}

Overall, the systems biology and spectral graph theory approach that we adopted in this study allowed us to identify putative proteins those can be termed as multi-cancer biomarkers.
In which, some proteins were already known to serve as candidate multi-cancer biomarkers that have confirmed the reliability of our results.
Our study has broadened the approach to identify cancer biomarkers using patterns corresponding to $-1$ eigenvalue degeneracy.
The selected five proteins {\it viz.,} {\it BMI}, {\it MAPK11}, {\it DDIT4}, {\it CDKN2A}, and {\it FYN} showed both biological significance and effectiveness in survival analysis.
The identification of multi-cancer biomarkers may lead to proposals of novel diagnostic tools and therapeutic schemes.
This finding could lead to another predictive angle and biological validation in the future.  
Furthermore, on technical ground, the article has presented a method to detect symmetrical patterns in weighted networks. 
The technique can be used to detect symetrical patterns in any networks generated from other real-world data. 


\section*{Additional files}
File of supplementary material comprises of Supplementary information, figures and tables. It lists information on Hallmarks of Cancer. It also includes table information on (Table S1) Datasets used for survival analysis. All datasets are considered for TCGA database, (Table S2) Datasets of seven cancers and their details, (Table S3) Top 10 degree nodes in weighted multi-cancer PPI network, (Table S4) Gene Expressions by risk groups, (Table S5) Biological functions of proteins, (Figure S1) Zero Degeneracy.

\section*{Availability of data and materials}
All data generated or analysed during this study are included in this article and its supplementary information files.
The software used in this paper to detect symmetrical patterns based on degenerate eigenvalues is available for download at https://github.com/ pramodsshinde/ networkSymmetry. All the codes were written in Matlab.

\section*{Acknowledgments}
\noindent We are grateful to unknown reviewers for helping us to improve the manuscript.
SJ thanks the support by grant of the ministry of education and science of the Russian Federation (Agreement No. 074-02-2018-330), Department of Science and Technology (DST), Government of India (EMR/2014/000368), Department of Atomic Energy, Government of India (37(3)/14/11/2018-BRNS/37131),  and Council of Scientific and Industrial Research (CSIR), Government of India (25(0293)/18/EMR-II).
PS thanks DST for the INSPIRE fellowship (IF150200). 
AY thanks CSIR, Government of India for the fellowship.
We acknowledge Dr. Hem Chandra Jha for interesting discussions. 
Authors thank Complex Systems Lab members for timely help and fruitful discussions. \\

\section*{Conflicts of interest}
Nothing to disclose.

\section*{Abbreviations}
PPI : Protein-protein interaction; OS: Overall survival; HR : Hazard ratio; BRCA : Breast cancer adenocarcinoma; 
CESC : Cervical squamous cell carcinoma; COAD : Colon adenocarcinoma; HNSC: Head and neck squamous cell carcinoma; LOAD : Lung adenocarcinoma; 
LUSC : Lung squamous cell carcinoma; OV : Ovarian serous cystadenocarcinoma; PRAD : Prostate adenocarcinoma

\label{lastpage}

\end{document}